\documentclass[a4paper,11pt]{article}
\usepackage{multirow}

\usepackage[english]{babel}
\usepackage{jinstpub}
\usepackage[sorting=none]{biblatex}
\usepackage[toc,page]{appendix}
\bibliography{references.bib}
\usepackage{graphicx}

\usepackage{float}
\title{A Precise Method to Determine the Energy Scale and Resolution using Gamma Calibration Sources in a Liquid Scintillator Detector}

\author[a]{Feiyang Zhang,}
\author[a]{Rui Li,}
\author[a]{Jiaqi Hui,}
\author[a,b]{Jianglai Liu,}
\author[a,1]{Yue Meng,\note{Corresponding author.}}
\author[a]{Yuanyuan Zhang}

\affiliation[a]{School of Physics and Astronomy, Shanghai Jiao Tong University, Shanghai, China}
\affiliation[b]{Tsung-Dao Lee Institute, Shanghai Jiao Tong University, Shanghai, China}

\emailAdd{mengyue@sjtu.edu.cn}

\abstract{
Gamma sources are routinely used to calibrate the energy scale and resolution of liquid scintillator detectors. However, non-scintillating material surrounding the source introduces energy losses, which may bias the determination of the centroid and width of the full absorption peak. 
In this paper, we present a general method to determine the true gamma centroid and width to a relative precision of 0.03\% and 0.50\%, respectively, using energy losses predicted by the Monte Carlo simulation. In particular, the accuracy of the assumed source geometry is readily obtained from the fit. The method performs well with experimental data in the Daya Bay detector.}

\keywords{Liquid scintillator, Calibration, Gamma sources, Full absorption peak, Compton tail, Energy scale, Energy resolution}

\begin{document}
\maketitle
\flushbottom

\section{Introduction}
\label{sec:intro}
For large liquid scintillator (LS) neutrino detectors, gamma sources are routinely deployed into various locations inside the detector to measure position-dependent energy responses~\cite{dayabay-calib-paper, Borexino-calib-paper, KamLAND-paper,double-chooz-calib-paper,RENO-paper}. 
For an ideal mono-energetic gamma peak, the visible energy shape can be approximated by a Gaussian function, with the centroid and width corresponding to the energy scale and resolution for that gamma energy. However, gamma sources are typically sealed with stainless steel enclosure, which causes energy losses primarily due to Compton scatterings, leading to a low energy tail which biases the fit. The situation becomes more delicate for the coming JUNO experiment, which has an unprecedented sub-percent requirement on the energy scale, and a 3\% requirement on the effective energy resolution~\cite{JUNO-calib-paper}.

A common practice is to truncate the Gaussian fit range towards lower energy. This reduces the tail contribution in the fit, but cannot remove it entirely. To fit the tail, approximate functions such as the Crystal Ball (CB) function~\cite{CB-paper} or Electromagnetic (EM) calorimeter function~\cite{EM-paper} have been adopted in the literature. 
For example, the EM calorimeter function assumes the tail shape follows a constant or an exponential distribution, which can fit gamma peaks as low as the 59.5~keV peak of $^{241}\rm Am$~\cite{Li-detector-paper}.
However, there are limitations to these methods since the tail shape is clearly geometry and energy dependent.

To properly take into account the expected leakage tail due to source geometry, in principle, the data can be compared to a complete Geant4 simulation\cite{dayabay-calib-paper, GT-calib-paper}. However, many detector and electronic parameters in the simulation have to be tuned to the data, which by itself can be a complex task. In this study, we propose a hybrid approach, in which the fit model is the deposited energy from the simulation, convolved with a simple Gaussian resolution function. Such an approach takes minimum simulation input, and the resolution function can be reliably obtained from the calibration data. Furthermore, geometrical uncertainty can be directly inferred from the fit, offering a strong check of the validity of the assumed geometry. 

To study this method, two JUNO prototype sources fabricated with well controlled geometry were deployed into a Daya Bay detector. In the rest of this paper, we shall discuss details of this method and its performance on the Daya Bay data, then quantify residual systematic uncertainties.

\section{Fitting model}
\begin{figure}
 \centering
 \includegraphics[width=4in]{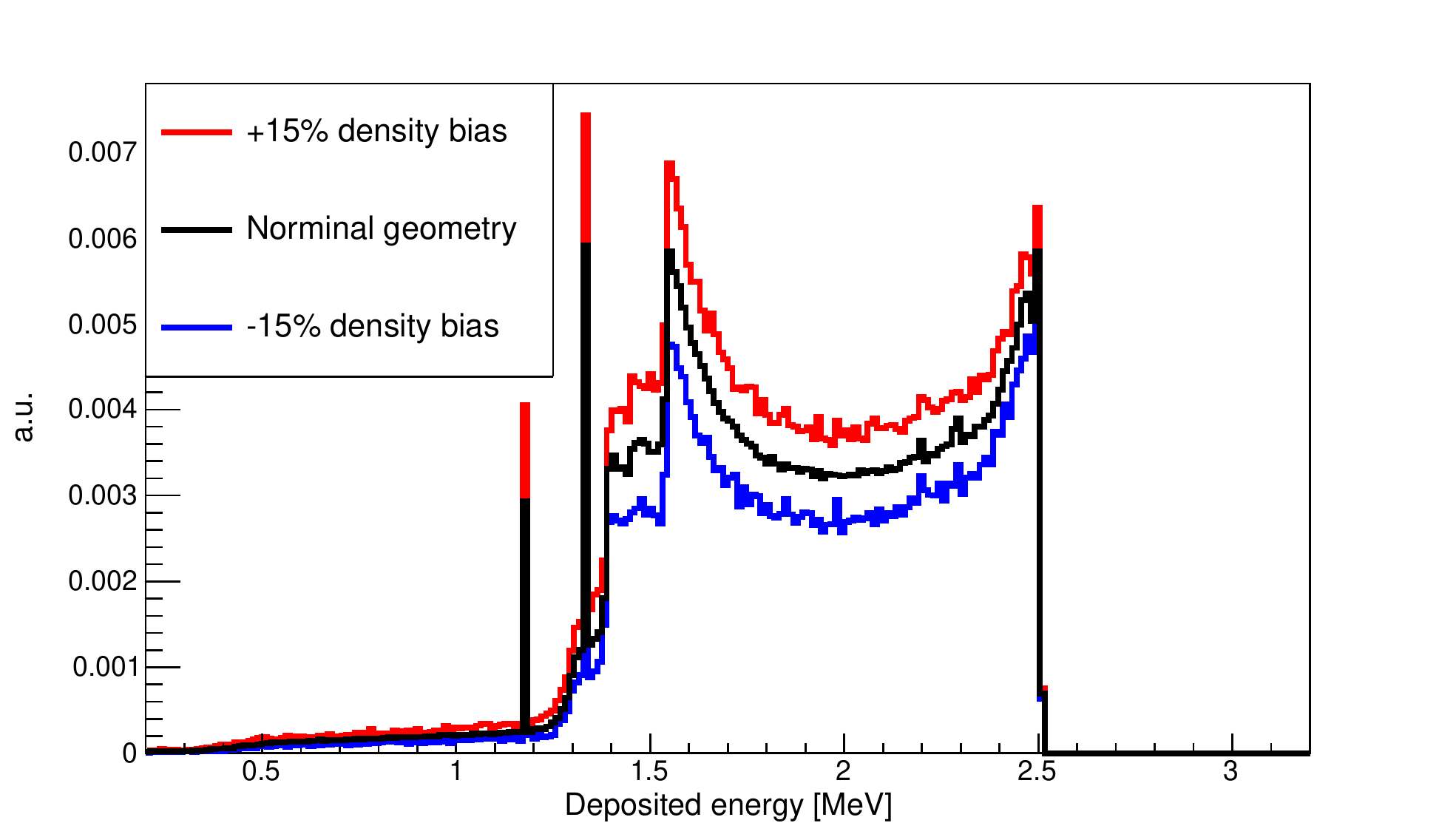}
 \caption{The distribution of the low energy tail of the deposited energy in the LS for a $^{60}\rm Co$ source. Also shown are the distributions when the density of the source enclosure is varied by $\pm$15\%. The spectra are binned into 1000 bins from 0.2 to 3.2~MeV. The number of events is normalized by the number of full absorption events.}
\label{fig:Co60_edep}
\end{figure}

Geant4~\cite{geant4-paper} is a mature tool to simulate deposited energy in a given material, based on 
the scattering cross section database~\cite{stopping-power-paper} 
and the photon transport model~\cite{photon-trans-paper}. For a $^{60}$Co source, with a cylindrical stainless steel enclosure of 6~mm in height and diameter~\cite{JUNO-calib-paper}, the deposited energy in the LS with the total absorption peak removed (or the "tail" in short) is shown in figure~\ref{fig:Co60_edep}. The uncertainty of the source geometry is studied by varying the density of the source enclosure by $\pm$15\%. The shape of the tail remains nearly the same, indicating that it is legitimate to use the nominal tail shape but floating its normalization in the fit model. 

The visible energy in the LS is defined as the total photoelectrons (PE) scaled by a constant value in MeV/PE. Its spectrum is a convolution of deposited energy with the detector resolution, symbolically described as:
\begin{equation}
    \label{equ:model}
    S_{\rm vis} = S_{\rm edep} \otimes {\rm resolution} \equiv S_{\rm peak} + S_{\rm tail}\,,
\end{equation}
separated into the peak and tail components. Specifically, the full absorption peak can be modeled as 
\begin{equation}
S_{\rm peak}(E;c, f, \sigma) = \frac{c}{\sigma \sqrt{2 \pi}} \cdot \exp(-\frac{(E- f E_{0})^{2}}{2 \sigma^{2}})\,,
\label{equ:peak:gaussian}
\end{equation}
in which a Gaussian resolution is assumed with $\sigma$ as the width at the peak, $E$ as the visible energy from the data, $E_{0}$ as the true energy of the gamma, and $c$ as a normalization constant. Note that a floating constant $f$ is taken to anchor the visible energy at $E_{0}$. The tail component can be modeled similarly as
\begin{equation}
    \label{equ:tail:gaussian}
    S_{\rm tail}(E;c,\beta, f, \sigma) = \sum_{i=1}^{N}P_{\rm tail}(E_{i})\cdot \frac{c\cdot\beta}{\sigma \sqrt{2\pi \cdot E_{i}/E_{0}}} \cdot \exp(-\frac{ (E - f E_{i})^{2}}{2\sigma^{2} \cdot E_{i}/E_{0}})\,,
\end{equation}
where $P_{\text {tail}}(E_{i})$ is the Probability Density Function (PDF) of the low energy tail from the simulation (figure.~\ref{fig:Co60_edep}), with $E_{i}$ as its ith energy bin. We have also taken the approximation that the visible energy is linear with deposited energy ($f E_i$) and that the detector resolution at this energy can be scaled by the factor $\sqrt{E_{i}/E_0}$. The tail takes the same normalization $c$ as the peak, but a floating parameter $\beta$ which allows a potential difference between the simulation and data. 

Note that more realistic resolution functions can replace the Gaussian if needed, e.g. a Poisson function if the photon statistics is low. Similarly, if the detector response is non-linear ($f$ is not a constant), 
or the resolution function does not scale with 
$\sqrt{1/E}$, 
more realistic functions can be easily incorporated into the tail (equation.~\ref{equ:tail:gaussian}).

\section{Fitting real data}
\label{dayabay:data}
\subsection{JUNO prototype sources in Daya Bay detector}
\begin{figure}
  \centering
  \includegraphics[width=2.05in]{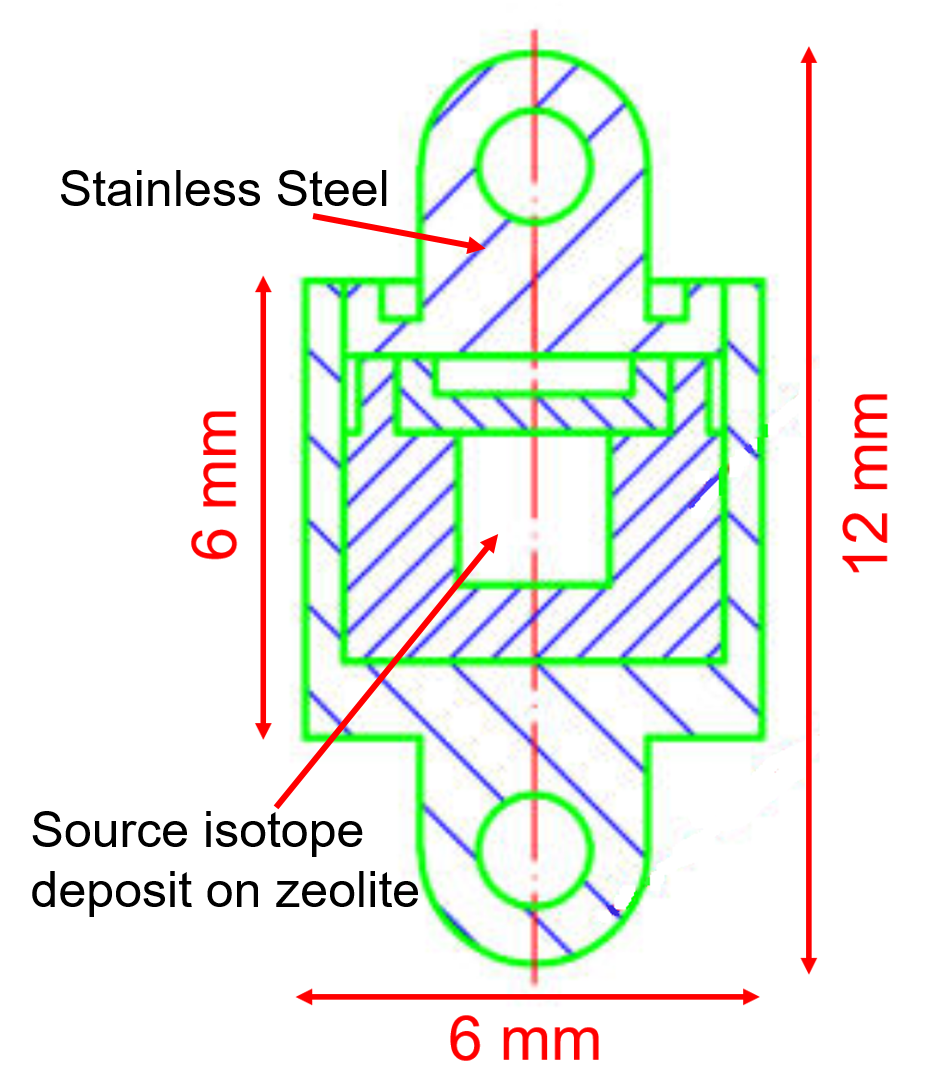}
  \includegraphics[width=1.90in]{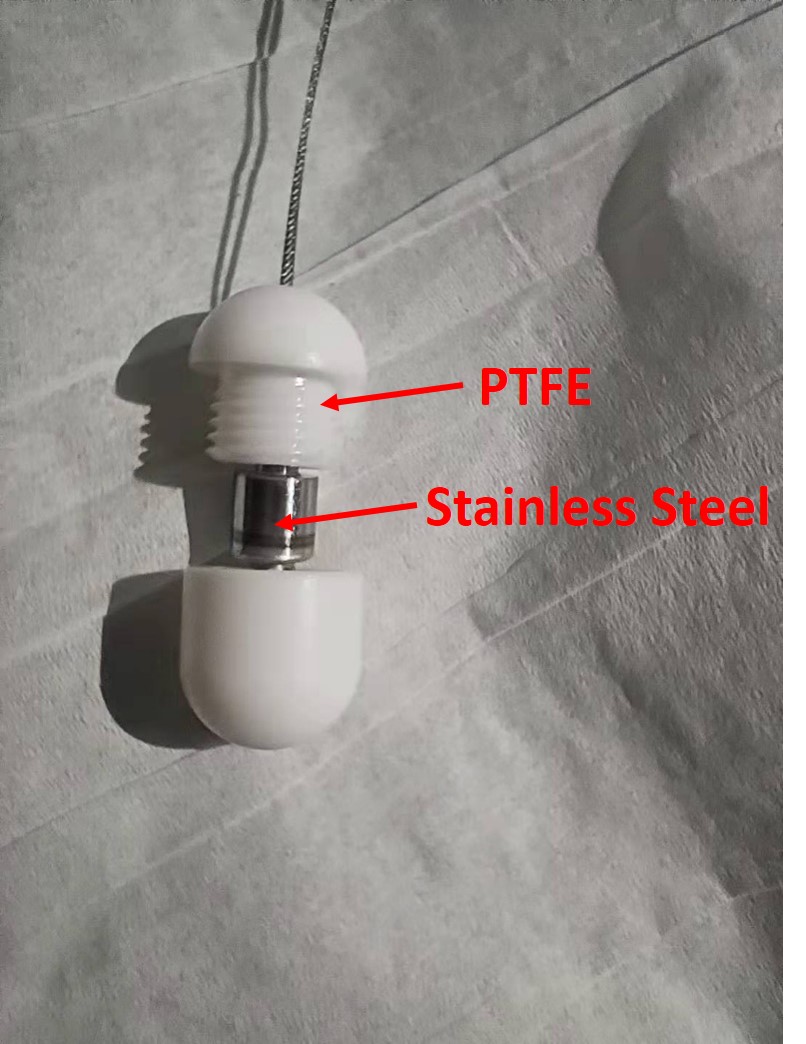}
  \includegraphics[width=1.85in]{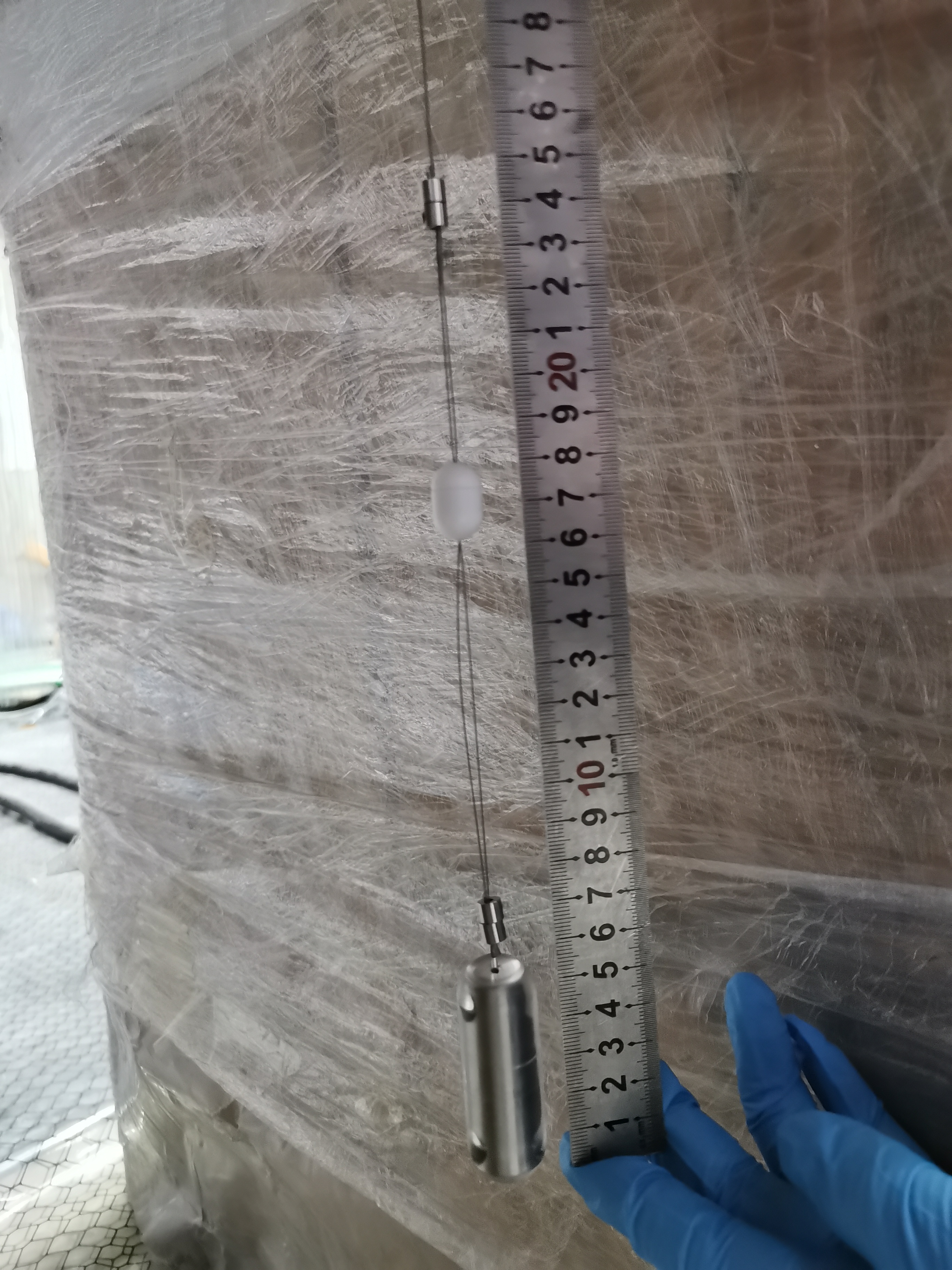}
    \caption{The $^{60}\rm Co$ source tested in the Daya Bay detector, left: source enclosure drawing, middle: source with PTFE reflector, right: the entire source assembly. A bottom weight of 50~g was used to maintain the tension of the cable.}
  \label{fig:source:enclosure}
\end{figure}

Two JUNO prototype sources $^{60}\rm Co$ and $^{137}\rm Cs$ ($\approx$~100~Bq) were fabricated at the vendor  
(Atomic Hi-tech~\cite{Atomic-Hi-tech-ref}), sealed with stainless steel enclosure as specified in figure~\ref{fig:source:enclosure}.
To test these sources, we deployed them into the center of a Daya Bay detector~\cite{dayabay-detector-paper} 
right after the end of its physics operation, with a position controlled to sub-cm precision~\cite{dayabay-ACU-paper}. As planned in JUNO, the radioactive source is further covered by a 
2~mm thick Polytetrafluoroethylene (PTFE) shell with much higher reflectivity than SS to minimize the optical photon loss. This PTFE layer also has some contribution to the energy loss tail. The entire source assembly is shown in figure~\ref{fig:source:enclosure}, including the cable, cable crimps, and the bottom weight.

For each source, the data taking time for the source-in run and source-out background run is 30~mins each. To avoid temporal drift of background (e.g., 609~keV $\gamma$ from radon decay), we have repeated such cycle every 3 hours several times and verified that the measured rate and spectrum are stable. A fiducial volume cut on the reconstructed radius from the detector $R$\textless1.2~m was applied to the data to suppress the ambient background. The true source spectrum is obtained by taking the difference between the source-in and source-out data. The measured visible energy spectra are shown in figure~\ref{fig:dayabay:exp:Co60} and figure~\ref{fig:dayabay:exp:Cs137} for $^{60}\rm Co$ and $^{137}\rm Cs$ respectively. 

\subsection{$^{60}\rm Co$ fits}

\begin{figure}
 \centering
 \includegraphics[width=2.9in]{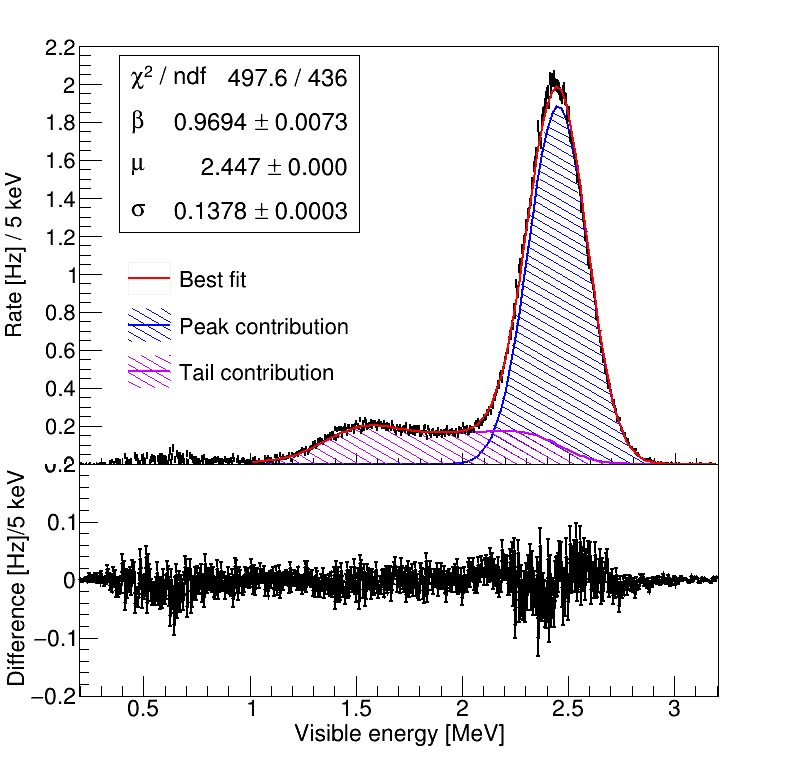}
 \includegraphics[width=2.9in]{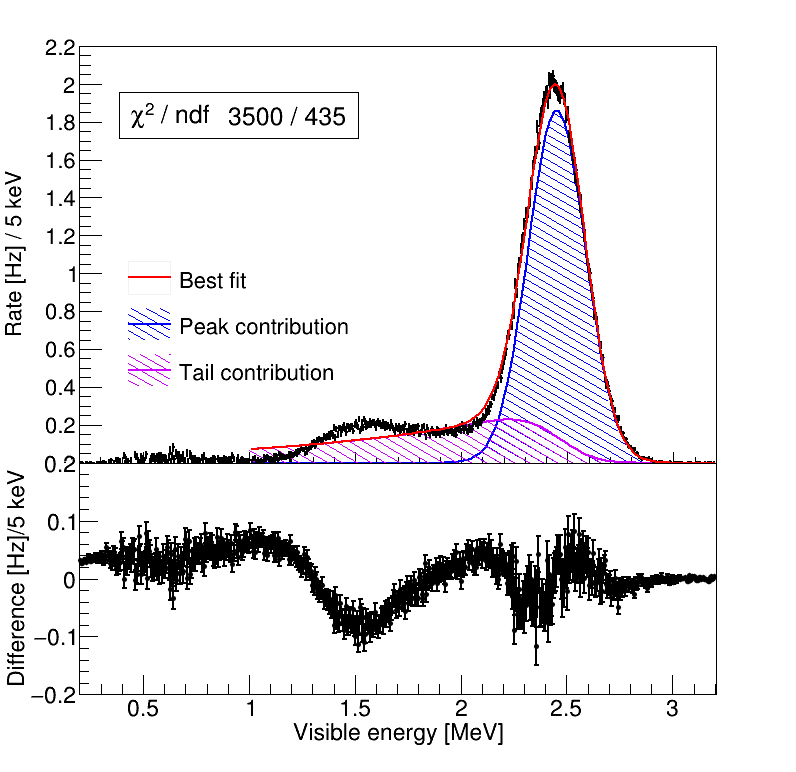}
 \caption{Visible energy of $^{60}\rm Co$ in the Daya Bay detector and the fits, left: with equation~\ref{equ:model}, and right: with EM calorimeter function. The bottom sub-figures are the difference between the best fit and data.}
 \label{fig:dayabay:exp:Co60}
\end{figure}

For the $^{60}\rm Co$ source, equation~\ref{equ:model} is applied to fit the data (left of figure~\ref{fig:dayabay:exp:Co60}). The fitting range is set from 1 to 3~MeV, covering both the peak and the tail. The fit quality is excellent. The fitted value of $\beta \approx$ 0.97, indicates that there is only 3\% difference between the measured and expected tail contributions. The 
EM calorimeter function fit is shown on the right of figure~\ref{fig:dayabay:exp:Co60} for comparison, which clearly cannot describe the complicated Compton tail shape. The uncertainty of the fit to the centroid and width of the full absorption peak shall be discussed in section~\ref{dayabay:sys:uncertinty}.

\subsection{$^{137}\rm Cs$ fit}
\begin{figure}
 \centering
 \includegraphics[width=3.0in]{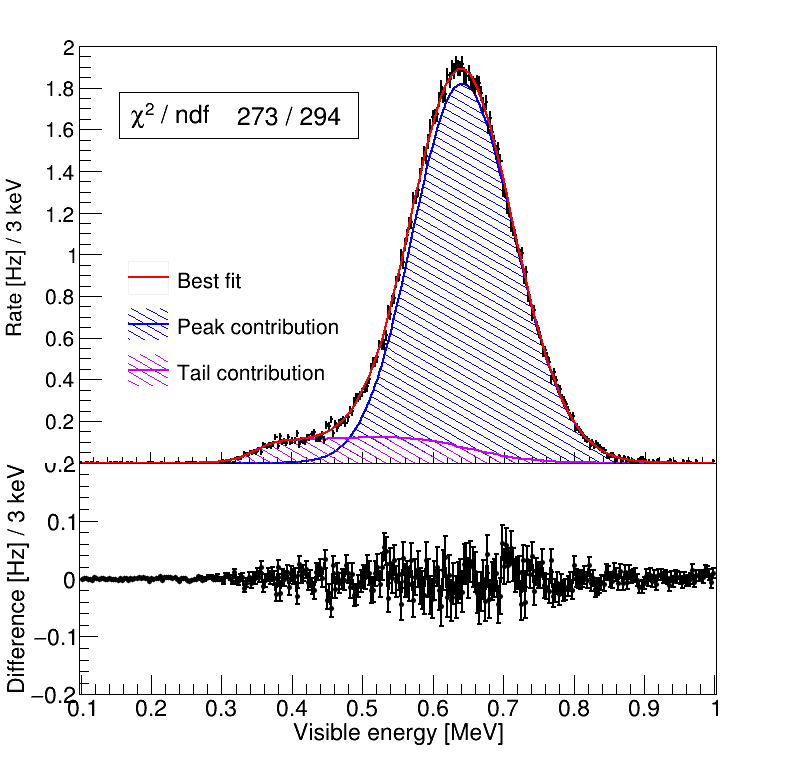}
 \caption{Visible energy of $^{137}\rm Cs$ in Daya Bay detector. The fit includes a threshold function and a Poisson resolution model.}
 \label{fig:dayabay:exp:Cs137}
\end{figure}

For $^{137}\rm Cs$, since the energy threshold of the Daya Bay detector is about 0.3~MeV~\cite{dayabay-eff-paper}, the trigger efficiency should be included in the fit. A Fermi-Dirac form, $1 /(e^{\frac{\mu - E}{w}}+1)$, is assumed for the efficiency, where $\mu$ is the threshold parameter, and $w$ is the width of the efficiency turning-on. The light yield of the Daya Bay detector is approximately 170~PE/MeV~\cite{dayabay-LY-paper}. Instead of using the Gaussian resolution model, the Poisson resolution model is adopted, which provides a better fit to the data. The result is shown in figure~\ref{fig:dayabay:exp:Cs137}, with a 
$\beta \approx$ 0.92. 
The extracted threshold value is 0.347$\pm$0.002~MeV, which is consistent with 0.37~MeV in ref.~\cite{dayabay-eff-paper}.

\section{Systematic uncertainties of the method}
\label{dayabay:sys:uncertinty}
Although in section~\ref{dayabay:data}, the fits perform quite well to the Daya Bay data, we need to estimate the projected systematic uncertainties when similar calibration sources are deployed into the JUNO detector.
We resort to simulations using JUNO parameters and make comparisons between the true input values and fitted values.
The gamma sources considered are: $^{137}\rm Cs$, $^{54}\rm Mn$, $^{40}\rm K$ and $^{60}\rm Co$, 
with geometry specified in ref.~\cite{JUNO-calib-paper}.
For each source, the PDF of the deposited energy in the LS is obtained via Geant4. These distributions are then folded with JUNO resolution functions to produce "measured" visible energy, on which fits are performed. The fit biases, defined as 
\begin{equation}
    \label{eq:bias:def}
    \rm Bias = \frac{\rm {fit\ result - true\ value}}{\rm {true\  value}}\times100\%\,,
\end{equation}
are categorized into three systematic components: 1) source geometry, 2) energy nonlinearity, and 3) energy resolution function.

\paragraph{Source geometry effect}
The deposited energy with imperfect knowledge of source geometry will obviously introduce a bias to the fit. 
As shown in section~\ref{dayabay:data}, our method introduces a $\beta$ parameter as an indicator of the geometry accuracy, 
which has been constrained to be better than 10\% for the two sources. To study potential biases, we take and smear the alternative energy deposition PDF with $\pm$15\% change of material density (blue and red histograms in figure~\ref{fig:Co60_edep}). We then fit them again by assuming the nominal PDF (black histogram in figure~\ref{fig:Co60_edep}). 
It is found that the mismatch on geometry can be mostly corrected by adjusting $\beta$ in the fit. The bias of the centroid and width are summarized in table~\ref{tab:sum:peak}.

\paragraph{Energy nonlinearity effect}
Equation~\ref{equ:model} assumes that the visible energy is linear to the deposited energy. In a real LS detector, physics and instrumental nonlinearities may be present, which alter the shape of the tail. To study the effect, 
the gamma nonlinearity in ref.~\cite{JUNO-calib-paper} (92\% to 101\% from 0.5~MeV to 3~MeV) is applied to $f$ in equation~\ref{equ:tail:gaussian} as the truth. The fit is repeated by maintaining a constant energy scale. The biases of the fits can be found in table~\ref{tab:sum:peak}. 

\paragraph{Energy resolution function effect}
Similarly, equation~\ref{equ:model} assumes that the smearing of the visible energy scales as 1/$\sqrt{E}$. More generally, non-stochastic smearing and PMT dark noises could introduce additional terms in the energy resolution function. To simulate a more realistic scenario, the energy resolution model in ref.~\cite{JUNO-calib-paper} is applied to the deposited energy, but the fit is performed by keeping the 1/$\sqrt{E}$ resolution function. The results are summarized in table~\ref{tab:sum:peak}. 

\begin{table}
  \centering
  \caption{Fit biases on the centroids and widths of different sources due to uncertainties in source geometry, energy nonlinearity, and energy resolution function. The numbers in the parentheses indicate uncertainties. See text for details.}
  \label{tab:sum:peak}
  \renewcommand\tabcolsep{2mm}
  \begin{tabular*}{150mm}{@{\extracolsep{\fill}}|c|ccc|ccc|}
   \hline
   \multirow{2}{*}{Source} & \multicolumn{3}{c|}{Centroid bias [\%]} & \multicolumn{3}{c|}{Width bias [\%]}  \\ 
   \cline{2-7}
   & Geometry & Nonlinearity & Resolution & Geometry & Nonlinearity & Resolution\\
   \hline
   $^{137}\rm Cs$& 0.021(2) & 0.005(2) & 0.000(2) & 0.46(4) & -0.10(4) & -0.01(4) \\
   $^{54}\rm Mn$ & 0.014(1) & 0.002(1) & 0.000(2) & 0.35(4) & -0.05(4) &  0.01(4) \\
   $^{40}\rm K$  & 0.005(1) & 0.000(1) & 0.000(1) & 0.18(4) & -0.01(4) &  0.00(4) \\
   $^{60}\rm Co$ & 0.012(1) & 0.001(1) & 0.000(1) & 0.50(4) & -0.02(4) &  0.02(4) \\
   \hline
  \end{tabular*}
\end{table}

One observes that our method is very robust to all three systematic effects. For all JUNO gamma sources considered here, the combined biases to the centroids and the widths of the full absorption peaks can be controlled to be better than 0.03\% and 0.50\%, respectively. Differences between the reality and our assumptions can be tuned further to improve. For example, the source geometry can be adjusted according to the $\beta$ parameter from the fit. The nonlinearity and resolution function can also be obtained from the source calibration and applied to the fit. 

\section{Summary}
In this paper, we develop a simple and robust method to fit calibration gamma peaks in a large LS detector, by properly taking into account the shape of the energy tail due to losses on the source enclosure. The method has been used to fit the measured spectra from two JUNO prototype sources in a Daya Bay detector, yielding excellent agreements between the data and fits. The systematic uncertainties to the centroids and widths of the gammas in JUNO have been estimated to be less than 0.03\% and 0.50\%. This method is extendable to determine the energy scale and resolution of a generic total absorption calorimeter.

\acknowledgments
This work is supported by the Strategic Priority Research Program of the Chinese Academy of Sciences, Grant No. XDA10010800, and the CAS Center for Excellence in Particle Physics (CCEPP). We thank the support from the Office of Science and Technology, Shanghai Municipal Government (Grant No. 16DZ2260200), and the support from the Key Laboratory for Particle Physics, Astrophysics and Cosmology, Ministry of Education.
We would like to thank the Daya Bay collaboration for providing the opportunity for our source measurements. We also thank the Daya Bay electronics and decommissioning teams for the local assistance.

\printbibliography

@article{JUNO-calib-paper,
    author = "Abusleme, Angel and others",
    collaboration = "JUNO",
    title = "{Calibration Strategy of the JUNO Experiment}",
    eprint = "2011.06405",
    archivePrefix = "arXiv",
    primaryClass = "physics.ins-det",
    doi = "10.1007/JHEP03(2021)004",
    journal = "JHEP",
    volume = "03",
    pages = "004",
    year = "2021"
}

@misc{stopping-power-paper,
  %author = {Martin Berger},
  %title = {ESTAR, PSTAR, and ASTAR: Computer Programs for Calculating Stopping-Power and Range Tables for Electrons, Protons, and Helium Ions},
  %year = {1992},
  %month = {1992-01-01},
  %publisher = {NIST Interagency/Internal Report (NISTIR), National Institute of Standards and Technology, Gaithersburg, MD},
  howpublished = {\url{https://www.physics.nist.gov/PhysRefData/Xcom/html/xcom1.html}}
  %language = {en},
}

@mastersthesis{CB-paper,
    author = "Oreglia, M.",
    title = "{A Study of the Reactions $\psi^\prime \to \gamma \gamma \psi$}",
    reportNumber = "SLAC-0236, SLAC-236, UMI-81-08973, SLAC-R-0236, SLAC-R-236",
    type = "Other thesis",
    month = "12",
    year = "1980"
}

@article{EM-paper,
    author = "Cheng, Jia-Hua and Wang, Zhe and Lebanowski, Logan and Lin, Guey-Lin and Chen, Shaomin",
    title = "{Determination of the total absorption peak in an electromagnetic calorimeter}",
    eprint = "1603.04433",
    archivePrefix = "arXiv",
    primaryClass = "physics.ins-det",
    doi = "10.1016/j.nima.2016.05.010",
    journal = "Nucl. Instrum. Meth. A",
    volume = "827",
    pages = "165--170",
    year = "2016"
}

@article{Li-detector-paper,
    author = "Rogers, F. and others",
    title = "{Large-area Si(Li) detectors for X-ray spectrometry and particle tracking in the GAPS experiment}",
    eprint = "1906.00054",
    archivePrefix = "arXiv",
    primaryClass = "physics.ins-det",
    doi = "10.1088/1748-0221/14/10/P10009",
    journal = "JINST",
    volume = "14",
    number = "10",
    pages = "P10009",
    year = "2019"
}

@article{GT-calib-paper,
    author = "Guo, Yuhang and Zhang, Qingmin and Zhang, Feiyang and Xiao, Mengjiao and Liu, Jianglai and Qu, Eryuan",
    title = "{Design of the Guide Tube Calibration System for the JUNO experiment}",
    eprint = "1905.02077",
    archivePrefix = "arXiv",
    primaryClass = "physics.ins-det",
    doi = "10.1088/1748-0221/14/09/T09005",
    journal = "JINST",
    volume = "14",
    number = "09",
    pages = "T09005",
    year = "2019"
}

@article{geant4-paper,
    author = "Agostinelli, S. and others",
    collaboration = "GEANT4",
    title = "{GEANT4--a simulation toolkit}",
    reportNumber = "SLAC-PUB-9350, FERMILAB-PUB-03-339, CERN-IT-2002-003",
    doi = "10.1016/S0168-9002(03)01368-8",
    journal = "Nucl. Instrum. Meth. A",
    volume = "506",
    pages = "250--303",
    year = "2003"
}

@article{dayabay-calib-paper,
    author = "Adey, D. and others",
    collaboration = "Daya Bay",
    title = "{A high precision calibration of the nonlinear energy response at Daya Bay}",
    eprint = "1902.08241",
    archivePrefix = "arXiv",
    primaryClass = "physics.ins-det",
    doi = "10.1016/j.nima.2019.06.031",
    journal = "Nucl. Instrum. Meth. A",
    volume = "940",
    pages = "230--242",
    year = "2019"
}

@article{dayabay-ACU-paper,
    author = "Liu, J. and others",
    title = "{Automated calibration system for a high-precision measurement of neutrino mixing angle $\theta_{13}$ with the Daya Bay antineutrino detectors}",
    eprint = "1305.2248",
    archivePrefix = "arXiv",
    primaryClass = "physics.ins-det",
    doi = "10.1016/j.nima.2014.02.049",
    journal = "Nucl. Instrum. Meth. A",
    volume = "750",
    pages = "19--37",
    year = "2014"
}

@inproceedings{KamLAND-paper,
    author = "Suekane, F. and Iwamoto, T. and Ogawa, H. and Tajima, O. and Watanabe, H.",
    collaboration = "KamLAND RCNS Group",
    title = "{An overview of the kamland 1-kiloton liquid scintillator}",
    booktitle = "{KEK - RCNP International School and Miniworkshop for Scintillating Crystals and their Applications in Particle and Nuclear Physics}",
    eprint = "physics/0404071",
    archivePrefix = "arXiv",
    month = "4",
    year = "2004"
}

@article{Borexino-calib-paper,
    author = "Caccianiga, Barbara and Re, Alessandra Carlotta",
    title = "{The calibration system for the Borexino experiment}",
    doi = "10.1142/S0217751X1442010X",
    journal = "Int. J. Mod. Phys. A",
    volume = "29",
    pages = "1442010",
    year = "2014"
}

@article{dayabay-eff-paper,
    author = "An, F. P. and others",
    collaboration = "Daya Bay",
    title = "{A side-by-side comparison of Daya Bay antineutrino detectors}",
    eprint = "1202.6181",
    archivePrefix = "arXiv",
    primaryClass = "physics.ins-det",
    doi = "10.1016/j.nima.2012.05.030",
    journal = "Nucl. Instrum. Meth. A",
    volume = "685",
    pages = "78--97",
    year = "2012"
}

@article{double-chooz-calib-paper,
title = {Double Chooz Calibration},
journal = {Nuclear Physics B - Proceedings Supplements},
volume = {229-232},
pages = {431},
year = {2012},
note = {Neutrino 2010},
issn = {0920-5632},
doi = {https://doi.org/10.1016/j.nuclphysbps.2012.09.068},
url = {https://www.sciencedirect.com/science/article/pii/S092056321200285X},
author = {Igor Ostrovskiy},
}

@article{RENO-paper,
    author = "KIM, Sang Yong",
    collaboration = "RENO",
    title = "{Prompt energy calibration at RENO}",
    doi = "10.1088/1742-6596/888/1/012136",
    journal = "J. Phys. Conf. Ser.",
    volume = "888",
    number = "1",
    pages = "012136",
    year = "2017"
}

@article{photon-trans-paper,
title = {A simple model of photon transport},
journal = {Nuclear Instruments and Methods in Physics Research Section B: Beam Interactions with Materials and Atoms},
volume = {101},
number = {4},
pages = {499-510},
year = {1995},
issn = {0168-583X},
doi = {https://doi.org/10.1016/0168-583X(95)00480-7},
url = {https://www.sciencedirect.com/science/article/pii/0168583X95004807},
author = {Dermott E. Cullen},
}

@article{dayabay-detector-paper, author = "An, F. P. and others", collaboration = "Daya Bay", title = "{The Detector System of The Daya Bay Reactor Neutrino Experiment}", eprint = "1508.03943", archivePrefix = "arXiv", primaryClass = "physics.ins-det", doi = "10.1016/j.nima.2015.11.144", journal = "Nucl. Instrum. Meth. A", volume = "811", pages = "133--161", year = "2016" }

@article{dayabay-LY-paper,
    author = "An, Feng Peng and others",
    collaboration = "Daya Bay",
    title = "{Measurement of electron antineutrino oscillation based on 1230 days of operation of the Daya Bay experiment}",
    eprint = "1610.04802",
    archivePrefix = "arXiv",
    primaryClass = "hep-ex",
    doi = "10.1103/PhysRevD.95.072006",
    journal = "Phys. Rev. D",
    volume = "95",
    number = "7",
    pages = "072006",
    year = "2017"
}

@misc{Atomic-Hi-tech-ref,
  title = {Atomic Hi-tech Company},
  howpublished = {\url{http://en.atom-hitech.com}}
}
\end{document}